\documentclass[10pt, a4paper]{article}

\usepackage[utf8]{inputenc}
\usepackage[T1]{fontenc}
\usepackage{lmodern,textcomp}
\usepackage[english]{babel}
\usepackage{csquotes}

\usepackage[top=3.9cm, bottom=4cm, left=3.5cm, right=3.5cm]{geometry}

\usepackage{eurosym,xspace}
\usepackage[nottoc]{tocbibind}
\usepackage{graphicx,subfig,float}
\usepackage[multiple]{footmisc}
\usepackage{authblk}
\usepackage{fancyhdr}
\usepackage[backend=bibtex8, citestyle=numeric-comp, maxbibnames=99, sorting=nyt,
			sortcites=true, firstinits=true, sorting=none]{biblatex}
\input{bibphys.def}

\usepackage{amsmath,amsfonts,amssymb}

\usepackage{maths_min}

\usepackage[pdftex, colorlinks=true, breaklinks=true,
	urlcolor=blue, linkcolor=blue, citecolor=red, linktocpage,
	pdfcreator={PDFLaTeX}, pdfauthor={Harold Erbin}
]{hyperref}
\usepackage[all]{hypcap}

\hypersetup{
	pdftitle={Supergravity, complex parameters and the Janis–Newman algorithm},
}

\numberwithin{equation}{section}

\newcommand{\email}[1]{\thanks{\href{mailto:#1}{\texttt{#1}}}}

\newcommand{\preprint}[0]{CPHT-001.0115}
\fancypagestyle{preprint}{
	\fancyhf{}

	\fancyhead[R]{\textsf{\preprint}}
}

\title{Supergravity, complex parameters and the Janis–Newman algorithm}

\author[1]{Harold Erbin\email{erbin@lpthe.jussieu.fr}}
\author[2]{Lucien Heurtier\email{lucien.heurtier@cpht.polytechnique.fr}}

\affil[1]{Sorbonne Universités, UPMC Univ Paris 06, UMR 7589, LPTHE, F-75005, Paris, France}
\affil[1]{CNRS, UMR 7589, LPTHE, F-75005, Paris, France}
\affil[2]{CPHT, École Polytechnique, CNRS, 91128 Palaiseau, France}

\begin{document}

\maketitle
\thispagestyle{preprint}

\begin{abstract}
The Demiański–Janis–Newman algorithm is an original solution generating technique.
For a long time it has been limited to producing rotating solutions, restricted to the case of a metric and real scalar fields, despite the fact that Demiański extended it to include more parameters such as a NUT charge.
Recently two independent prescriptions have been given for extending the algorithm to gauge fields and thus electrically charged configurations.
In this paper we aim to end setting up the algorithm by providing a missing but important piece, which is how the transformation is applied to complex scalar fields.
We illustrate our proposal through several examples taken from $N = 2$ supergravity, including the stationary BPS solutions from Behrndt et al. and Sen's axion–dilaton rotating black hole.
Moreover we discuss solutions that include pairs of complex parameters, such as the mass and the NUT charge, or the electric and magnetic charges, and we explain how to perform the algorithm in this context (with the example of Kerr–Newman–Taub–NUT and dyonic Kerr–Newman black holes).
The final formulation of the DJN algorithm can possibly handle solutions with five of the six Plebański–Demiański parameters along with any type of bosonic fields with spin less than two (exemplified with the SWIP solutions).
This provides all the necessary tools for applications to general matter-coupled gravity and to (gauged) supergravity.
\end{abstract}

\tableofcontents

\section{Introduction}

The Janis–Newman (JN) algorithm is a peculiar solution generating technique.
It was originally designed to add an angular momentum $a$ to a static solution with mass $m$ and electric charge $q$~\cite{newman_note_1965, newman_metric_1965}, before being extended by Demiański and Newman to add a NUT charge $n$~\cite{demianski_combined_1966, demianski_new_1972} – we will call this version the Demiański–Janis–Newman (DJN) algorithm.
Recent reviews of the JN algorithm can be found in~\cite{adamo_kerr-newman_2014, erbin_deciphering_2014}.

Supergravity rotating solutions is an intense field of research, and it is surprising that the (D)JN algorithm has almost never been applied in this context (with the exception of~\cite{yazadjiev_newman-janis_2000}).
One explanation is that such theories present a number of gauge fields and complex scalar fields that could not be transformed in the original formulation of the DJN algorithm.
For instance, Yazadjiev~\cite{yazadjiev_newman-janis_2000} showed that it was possible to obtain the metric and the dilaton of Sen's dilaton–axion charged rotating black hole~\cite{sen_rotating_1992} (non-extremal solution of the $T^3$ model), but did not succeed in finding the axion nor the gauge field.

Each of these problems possess a different explanation.
First of all, it was not known how to perform the transformation on the gauge field until recently, where two different prescriptions have been proposed~\cite{keane_extension_2014, erbin_janis-newman_2015, adamo_kerr-newman_2014}.

The second problem can be traced to the fact that the dilaton and the axion are naturally gathered into a complex scalar field, and any attempt to transform each field independently can only fail, the reason being that the axion is vanishing for the static configuration, while it is non-zero for the rotating black hole.
Moreover the usual transformation rules can not be applied to complex scalar fields because they include a reality condition which is a too strong requirement for transforming complex fields, and one of the goal of the paper is to show how to modify the original prescription to accommodate this new fact.
We will illustrate this proposal on several examples, all taken from $N = 2$ ungauged supergravity, completing Yazadjiev's analysis~\cite{yazadjiev_newman-janis_2000} of Sen's rotating black hole, and showing how some BPS rotating black holes from~\cite{behrndt_stationary_1998} can be obtained (which includes solutions from pure supergravity and from the $STU$ model).

Another issue arises when one considers the NUT charge.
Indeed a long-standing problem of the DJN algorithm was the impossibility to find the metric function using the usual rules of the JN algorithm.
We recently demonstrated~\cite{erbin_deciphering_2014} how to extend the algorithm by complexifying also the mass $m = m' + i n$, and we will recall the details with the example of the Kerr–Newman–Taub–NUT solution.

A related case is the dyonic Reissner–Nordström with electric and magnetic charges $q$ and $p$, which can be used as a seed metric for deriving the dyonic Kerr–Newman solution.
It is necessary to follow the recipe of the previous examples, since the original JN rules are failing again.
This is related to the fact that the electric and magnetic charges are naturally associated into the (complex) central charge $Z = q + i p$.
In this way we succeed in performing the JN algorithm to a solution with magnetic charges.

Parameters are naturally gathered into complex pairs in the framework of Plebański--Demiański formalism~\cite{plebanski_class_1975, plebanski_rotating_1976} and thus it should not be too surprising to find again that these combinations are the one which should be used for a consistent DJN algorithm.
Moreover the fact that coordinates and parameters can be complexified for generating new solutions was already present in the formalism of Quevedo~\cite{quevedo_complex_1992, quevedo_determination_1992}.
Finally let's note that all these solutions can be embedded into $N = 2$ supergravity~\cite{alonso-alberca_supersymmetry_2000, klemm_supersymmetry_2013}.

The paper is organized as follows.
In section~\ref{sec:janis-newman} we recall the DJN algorithm tools derived in ~\cite{erbin_deciphering_2014}, presenting its main steps and formulas.
In sections~\ref{sec:taub-nut} and~\ref{sec:dyonic-kerr} we show how to obtain the Kerr–Newman–Taub-NUT and dyonic Kerr–Newman solutions using the latter.
Finally in section~\ref{sec:N=2-ungauged-sugra} we explain how to extend the JN algorithm to complex scalar fields in the context of $N = 2$ supergravity.
Appendix~\ref{app:sugra-nut} contains a discussion on the addition of a NUT charge for solutions of pure $N = 2$ supergravity, along with a discussion of SWIP solutions.

\section{Janis-Newman algorithm}
\label{sec:janis-newman}

In this section we summarize the main idea of the Demiański–Janis–Newman algorithm~\cite{newman_note_1965, demianski_new_1972} and we give general formulas derived in~\cite{erbin_deciphering_2014}.
While the original algorithm has been formulated in terms of Newman–Penrose formalism, we will use the simpler prescription due to G. Giampieri, which has been shown to be equivalent~\cite{giampieri_introducing_1990, erbin_janis-newman_2015}.
For details on the first formulation we refer the reader to the literature~\cite{newman_note_1965, newman_metric_1965, drake_uniqueness_2000, keane_extension_2014, adamo_kerr-newman_2014}.

Let's consider the metric and gauge field
\begin{subequations}
\begin{gather}
	\label{eq:static-metric}
	\dd s^2 = - f_t(r)\, \dd t^2 + f_r(r)\, \dd r^2 + f_\Omega(r)\, \dd\Omega^2, \\
	\label{eq:static-gauge-field}
	A(r) = f_A(r)\, \dd t,
\end{gather}
\end{subequations}
and a set of real scalar fields $\chi_a(r)$, where
\begin{equation}
	\dd \Omega^2 = \dd\theta^2 + H(\theta)^2\, \dd \phi^2, \qquad
	H(\theta) \equiv \sin \theta
\end{equation} 
($H$ being used for shortening subsequent expressions).
It is not necessary to specify the action for performing the algorithm as one needs only the expressions of the various seed fields, but one needs to check that the result is a solution of the equations of motion.
Indeed it is not fully understood under which conditions the algorithm will send a solution to another solution (for some proofs and discussions, see~\cite{talbot_newman-penrose_1969, demianski_new_1972, schiffer_kerr_1973, gurses_lorentz_1975, drake_uniqueness_2000, pirogov_towards_2013, hansen_applicability_2013, erbin_deciphering_2014, adamo_kerr-newman_2014}).
In this paper we restrict ourselves to vanishing cosmological constant.

The DJN algorithm requires introducing the null coordinates
\begin{equation}
	\label{change:null-diff}
	\dd t = \dd u + \sqrt{\frac{f_r}{f_t}}\, \dd r,
\end{equation} 
and the metric and gauge field resulting from \eqref{eq:static-metric} and \eqref{eq:static-gauge-field} are
\begin{subequations}
\begin{gather}
	\label{eq:static-metric:ur}
	\dd s^2 = - f_t\, \dd u^2 - 2 \sqrt{f_t f_r}\; \dd u \dd r + f_\Omega(r)\, \dd\Omega^2, \\
	\label{eq:static-gauge-field:ur}
	A = f_A\, \dd u,
\end{gather}
\end{subequations}
where the $r$-component of the gauge field has been removed with a gauge transformation, a step which is primordial for having a consistent DJN transformation.

The DJN starts by letting the coordinates $u$ and $r$ to be complex under the condition that the metric, the gauge fields and the scalar fields are still real.

Then one can perform the complex change of variables
\begin{equation}
	\label{change:jna-coord}
	r = r' + i\, F(\theta), \qquad
	u = u' + i\, G(\theta),
\end{equation} 
where $u', r' \in \R$ and~\footnote{An extra parameter can be added~\cite{demianski_new_1972, erbin_deciphering_2014}, but we will not need it in this work.}
\begin{equation}
	F(\theta) = n - a\, H'(\theta), \qquad
	G(\theta) = a\, H'(\theta) - 2 n \ln H(\theta),
\end{equation} 
or by replacing $H = \sin \theta$
\begin{equation}
	\label{eq:jna-functions}
	F(\theta) = n - a \cos \theta, \qquad
	G(\theta) = a \cos \theta - 2 n \ln \sin \theta.
\end{equation} 
The parameters $a$ and $n$ are respectively interpreted as the angular momentum and the NUT charge of the solution.

The differentials of \eqref{change:jna-coord}
\begin{equation}
	\dd r = \dd r' + i\, F'\, \dd\theta, \qquad
	\dd u = \dd u' + i\, G'\, \dd\theta
\end{equation} 
are complex and they would spoil the reality of the metric.
For this reason one makes the ansatz
\begin{equation}
	i\, \dd \theta = H\, \dd \phi
\end{equation} 
resulting into
\begin{equation}
	\label{change:jna-diff}
	\dd r = \dd r' + F' H\, \dd \phi, \qquad
	\dd u = \dd u' + G' H\, \dd \phi.
\end{equation} 
The form of the ansatz is justified from comparison with the tetrad formalism~\cite{giampieri_introducing_1990, erbin_janis-newman_2015, ferraro_untangling_2014}.
For comprehensiveness the derivatives of \eqref{eq:jna-functions} are given
\begin{equation}
	\label{eq:jna-functions-deriv}
	F'(\theta) = a \sin \theta, \qquad
	G'(\theta) = - a \sin \theta - 2 n \tan \theta.
\end{equation} 

Along these coordinates transformations, the four $r$-dependent functions and the scalar fields
\begin{equation}
	\mc F_i(r) \equiv \{ f_t, f_r, f_\Omega, f_A, \chi_a \}
\end{equation} 
transform accordingly into
\begin{equation}
	\widetilde{\mc F_i} \equiv \{ \tilde f_t, \tilde f_r, \tilde f_\Omega, \tilde f_A, \tilde \chi_a \}.
\end{equation} 
We will refer to this step as the "complexification" of the functions $\mc F_i$ even if the functions $\widetilde{\mc F_i}$ are real.
There are only two conditions that we impose on these functions
\begin{equation}
	\widetilde{\mc F_i} = \widetilde{\mc F_i} \big(r', F(\theta) \big) \in \R, \qquad
	\widetilde{\mc F_i}(r, 0) = \mc F_i(r).
\end{equation} 
Note that the primes will be often omitted once transformations on the coordinates are performed in the rest of the paper. 
The way to transform the functions is not constrained by the algorithm itself and the choice is somewhat arbitrary.
Nonetheless a set of rules have been found from various examples
\begin{subequations}
\label{eq:complexification-rules}
\begin{align}
	\label{eq:complexification-rules-r}
	r & \longrightarrow \frac{1}{2} (r + \bar r) = \Re r\,, \\
	\label{eq:complexification-rules-1/r}
	\frac{1}{r} & \longrightarrow \frac{1}{2} \left(\frac{1}{r} + \frac{1}{\bar r}\right) = \frac{\Re r}{\abs{r}^2}\,, \\
	\label{eq:complexification-rules-r2}
	r^2 & \longrightarrow \abs{r}^2,
\end{align}
\end{subequations}
and it was explained in~\cite[app.~B]{erbin_janis-newman_2015} that several of these possibilities are perfectly equivalent.
As we will see in the rest of this paper, these rules do not apply when $n \neq 0$ or in presence of complex scalar fields.

Performing the DJN transformations with \eqref{change:jna-coord} and \eqref{change:jna-diff} on the metric \eqref{eq:static-metric:ur} and gauge field \eqref{eq:static-gauge-field:ur} gives
\begin{subequations}
\begin{gather}
	\dd s^2 = - \tilde f_t (\dd u + \alpha\, \dd r + \omega H\, \dd\phi )^2
		+ 2 \beta\, \dd r \dd \phi
		+ \tilde f_\Omega (\dd\theta^2 + \sigma^2 H^2 \dd\phi^2), \\
	A = \tilde f_A\, (\dd u + G' H\, \dd \phi)
\end{gather}
\end{subequations}
where the following quantities have been defined
\begin{equation}
	\omega = G' + \sqrt{\frac{\tilde f_r}{\tilde f_t}}\, F', \qquad
	\sigma^2 = 1 + \frac{\tilde f_r}{\tilde f_\Omega}\, F'^2, \qquad
	\alpha = \sqrt{\frac{\tilde f_r}{\tilde f_t}}, \qquad
	\beta = \tilde f_r\, F' H.
\end{equation}

Finally Boyer–Lindquist (BL) coordinates follow from the transformation
\begin{equation}
	\label{change:bl-diff}
	\dd u = \dd t - g(r) \dd r, \qquad
	\dd \phi = \dd \phi' - h(r) \dd r
\end{equation} 
with
\begin{equation}
	\label{change:bl-gh}
	g(r) = \frac{\sqrt{\big(\tilde f_t \tilde f_r \big)^{-1}}\, \tilde f_\Omega - F' G'}{\Delta}, \qquad
	h(r) = \frac{F'}{H(\theta) \Delta}, \qquad
	\Delta = \frac{\tilde f_\Omega}{\tilde f_r} + F'^2
		= \frac{\tilde f_\Omega}{\tilde f_r}\, \sigma^2.
\end{equation} 
Let us insist on the fact that the transformation is well defined only if $g$ and $h$ are independent of $\theta$ when plugging the explicit expressions of the functions.

Finally the metric in $(t, r)$ coordinates can be written
\begin{subequations}
\label{eq:stationary-metric-gauge-field:tr}
\begin{gather}
	\label{eq:stationary-metric:tr}
	\dd s^2 = - \tilde f_t (\dd t + \omega H\, \dd\phi )^2
		+ \tilde f_\Omega \left( \frac{\dd r^2}{\Delta} + \dd\theta^2 + \sigma^2 H^2 \dd\phi^2 \right), \\
	\label{eq:stationary-gauge-field:tr}
	A = \tilde f_A\, \left( \dd t - \frac{\tilde f_\Omega}{\sqrt{\tilde f_t \tilde f_r}\, \Delta}\, \dd r + G' H\, \dd \phi \right).
\end{gather}
\end{subequations}
In most of the cases $A_r$ will depend only on $r$ and can consequently be removed by a gauge transformation.

\section{Kerr–Newman–Taub–NUT solution}
\label{sec:taub-nut}

A long-standing difficulty of Demiański's extension of the JN algorithm~\cite{demianski_new_1972} was the impossibility to find the complexification of the metric function that was leading from Schwarzschild to Kerr–Taub–NUT.
In this section we recall the solution to this problem that we gave in a previous paper~\cite{erbin_janis-newman_2015}, where we extended Demiański's result to Kerr–Newman–Taub–NUT.

Reissner–Nordström metric is given by
\begin{subequations}
\label{eq:reissner-nordstrom}
\begin{equation}
	\label{metric:reissner-nordstrom:tr}
	\dd s^2 = - f(r)\, \dd t^2 + f(r)^{-1}\, \dd r^2 + r^2 \dd \Omega^2, \qquad
	f(r) = 1 - \frac{2m}{r} + \frac{q^2}{r^2},
\end{equation} 
$m$ and $q$ being the mass and the electric charge, and the electromagnetic gauge field reads
\begin{equation}
	\label{pot:reissner-nordstrom:tr}
	A = \frac{q}{r}\; \dd t.
\end{equation} 
\end{subequations}

As explained above, applying the algorithm described in section~\ref{sec:janis-newman} does not lead to the KN–TN solution.
In fact one needs to also complexify the \emph{mass}.
In this case the function $f$ is complexified as
\begin{equation}
	\tilde f = 1 - \left(\frac{m}{r} + \frac{\bar m}{\bar r} \right) + \frac{q^2}{\abs{r}^2}
		= 1 - \frac{2 \Re(m \bar r) + q^2}{\abs{r}^2},
\end{equation} 
and performing the transformation
\begin{equation}
	\label{eq:complex-mass-no-lambda}
	m = m' + i n, \qquad
	r = r' + i F
\end{equation} 
gives (omitting the primes)
\begin{equation}
	\tilde f = 1 - \frac{2 m r + 2 n F}{\rho^2}, \qquad
	\rho^2 = r^2 + F^2.
\end{equation}

Considering the transformations \eqref{eq:jna-functions} leads to
\begin{equation}
	\tilde f = 1 - \frac{2 m r - q^2 + n (n - a \cos \theta)}{\rho^2}, \qquad
	\rho^2 = r^2 + (n - a \cos \theta)^2.
\end{equation} 
The metric and the gauge fields in BL coordinates can be read from \eqref{eq:stationary-metric-gauge-field:tr} to be
\begin{subequations}
\begin{gather}
	\dd s^2 = - \tilde f\, (\dd t + \Omega\, \dd\phi )^2
		+ \frac{\rho^2}{\Delta}\, \dd r^2
		+ \rho^2 (\dd\theta^2 + \sigma^2 H^2 \dd\phi^2), \\
	A = \frac{q}{\rho^2}\, \Big( \dd t - (a \sin^2 \theta + 2 n \cos \theta) \dd \phi \Big) + A_r\, \dd r.
\end{gather}
\end{subequations}
One can check that $A_r$ is a function of $r$ only
\begin{equation}
	A_r = - \frac{q}{\Delta}
\end{equation} 
and it can be removed by a gauge transformation.
The various quantities that appear are given by
\begin{equation}
	\Omega = - 2 n \cos \theta - (1 - \tilde f^{-1})\, a \sin^2 \theta, \qquad
	\sigma^2 = \frac{\Delta}{\tilde f \rho^2}, \qquad
	\Delta = \tilde f \rho^2 + a^2 \sin^2 \theta.
\end{equation} 
This corresponds to the Kerr–Newman–Taub–NUT solution~\cite{alonso-alberca_supersymmetry_2000}.

\section{Dyonic Kerr–Newman black hole}
\label{sec:dyonic-kerr}

The dyonic Reissner–Nordström metric is obtained from the electric one \eqref{eq:reissner-nordstrom} by the replacement~\cite[sec.~6.6]{carroll_spacetime_2004}
\begin{equation}
	q^2 \longrightarrow \abs{Z}^2 = q^2 + p^2
\end{equation} 
and the transformation is unchanged.
The symbol $Z$ corresponds to the central charge~\cite{alonso-alberca_supersymmetry_2000}
\begin{equation}
	Z = q + i p.
\end{equation} 
This is particularly useful when looking at the dyonic RN as a solution of pure $N = 2$ ungauged supergravity.
Then the metric function reads
\begin{equation}
	f(r) = 1 - \frac{2m}{r} + \frac{\abs{Z}^2}{r^2}.
\end{equation} 
On the other hand the gauge field receives a new contribution and one has~\cite{alonso-alberca_supersymmetry_2000}
\begin{equation}
	A = \frac{q}{r}\, \dd t + p \cos \theta\, \dd\phi
		= \frac{q}{r}\, \dd u + p \cos \theta\, \dd\phi
\end{equation}
(the last equality arising after a gauge transformation).

For simplifying the computations we only consider the case $n = 0$ with
\begin{equation}
	F = - a\, \cos \theta, \qquad
	G = a\, \cos \theta,
\end{equation} 
but the general case $n \neq 0$ follows directly.
The transformation of the metric is totally identical to the previous case (section~\ref{sec:taub-nut}) and one needs only to focus on the gauge field.

One has to rewrite first the gauge field as
\begin{equation}
	A = \Re\left(\frac{Z}{r}\right)\, \dd t + p \cos \theta\, \dd\phi
\end{equation} 
before performing the JN transformation.
The first term is complexified as
\begin{equation}
	\Re\left(\frac{Z}{r}\right) = \Re\left(\frac{Z r}{r^2}\right)
		= \frac{\Re(Z r)}{\abs{r}^2}
\end{equation} 
and inserting the transformation
\begin{equation}
	r = r' + i a \cos \theta
\end{equation} 
gives
\begin{equation}
	A = \frac{q r - p a \cos \theta}{\rho^2}\, (\dd u - a \sin^2 \theta\, \dd\phi ) + p \cos \theta \, \dd\phi.
\end{equation} 

After changing coordinates into the BL system, the $A_r$ term is
\begin{equation}
	\Delta\, A_r = - \frac{q r - p a \cos \theta}{\rho^2}\, \rho^2 - p a \cos \theta = - q r
\end{equation} 
($\Delta(r)$ is the denominator of the BL functions, not the Laplacian).
Since $A_r = A_r(r)$ one can remove it and obtains finally
\begin{equation}
	A = \frac{q r - p a \cos \theta}{\rho^2}\, (\dd t - a \sin^2 \theta\, \dd\phi ) + p \cos \theta \, \dd\phi.
\end{equation} 
Using the fact that
\begin{equation}
	a^2 \sin^2 \theta = r^2 + a^2 - \rho^2
\end{equation} 
we rewrite it
\begin{subequations}
\begin{align}
	A &= \frac{q r}{\rho^2} (\dd t - a \sin^2 \theta \dd\phi)
			+ \frac{p \cos \theta}{\rho^2} \left(a\, \dd t + (r^2 + a^2)\, \dd\phi \right) \\
		&= \frac{q r - p a \cos \theta}{\rho^2}\, \dd t
			+ \left(- \frac{q r}{\rho^2}\, a \sin^2 \theta + \frac{p(r^2 + a^2)}{\rho^2}\, \cos\theta \right) \dd\phi\,,
\end{align}
\end{subequations}
as it is presented in~\cites{alonso-alberca_supersymmetry_2000}[sec.~6.6]{carroll_spacetime_2004}.

The Yang–Mills Kerr–Newman black hole found by Perry~\cite{perry_black_1977} can also be derived in this way.

\section{Janis–Newman algorithm for complex fields}
\label{sec:N=2-ungauged-sugra}

In this section we expose the main ingredient for applying the JN transformation with $a \neq 0$ (but $n = 0$) on complex scalar fields, which is that one needs to transform together the real and imaginary parts without enforcing any reality condition.
Solutions with $n \neq 0$ require a more careful treatment and are studied in appendix~\ref{app:sugra-nut}.

We will give examples from ungauged $N = 2$ supergravity coupled to $n_v = 0, 1, 3$ vector multiplets (pure supergravity, $STU$ model and $T^3$ model).
Our aim is not to give a detailed account of supergravity, and the interested reader may look at usual references~\cite{freedman_supergravity_2012, andrianopoli_general_1996, andrianopoli_n2_1997}.

\subsection{Rule for complex fields}
\label{sec:N=2-ungauged-sugra:complex-rule}

Let's consider a complex scalar field $\chi$ such that
\begin{equation}
	\chi(r) = 1 + \frac{R}{r}
\end{equation} 
for the static configuration, $R$ being a parameter.
This is a very typical behaviour, where the imaginary part vanishes and the real part is harmonic with respect to the $3$-dimensional spatial metric.

The first step of the JN algorithm is to complexify all the fields, using only the fact that $r$ is complex.
Namely, performing the JN transformation \eqref{change:jna-coord}
\begin{equation}
	r = r' - i a \cos \theta
\end{equation} 
gives
\begin{equation}
	\label{eq:scalar-complexification-rule}
	\tilde \chi
		= 1 + \frac{R}{r' - i a \cos \theta}
		= 1 + \frac{R\, (r' + i a \cos \theta)}{\rho^2},
\end{equation}
where as usual $\rho^2 = r'^2 + a^2 \cos^2 \theta$.

The imaginary part is thus proportional to the angular momentum $a$.
Consequently it is impossible to generate the latter only from the static imaginary part since the traditional JN algorithm can not generate a non-zero rotating field from a null static one.
The main argument for this new rule is that one should not enforce any reality condition on the real or imaginary parts because they naturally form a pair.
In other words, imaginary and real parts of the scalar fields naturally form a pair which cannot be reduced by any reality condition.
Splitting a complex fields into its real and imaginary parts may hence obscure its structure and leads to a failure of the transformation (as it shows up in~\cite{yazadjiev_newman-janis_2000}).
Note also that $\tilde \chi$ is now a complex harmonic function.

\subsection{Review of \texorpdfstring{$N=2$}{N = 2} ungauged supergravity}

The gravity multiplet contains the metric and the graviphoton
\begin{equation}
	\{ g_{\mu\nu}, A^0 \}
\end{equation} 
while each of the vector multiplets contains a gauge field and a complex scalar field
\begin{equation}
	\{ A^i, z^i \}, \qquad i = 1, \ldots, n_v.
\end{equation} 
The scalar fields $z^i$ (we denote the conjugate fields by $\bar z^i = z^{\bar\imath}$) parametrize a special Kähler manifold with metric $g_{i\bar\jmath}$.
This manifold is uniquely determined by an holomorphic function called the prepotential $F$.
The latter is better defined using the homogeneous (or projective) coordinates $X^\Lambda$ such that
\begin{equation}
	z^i = \frac{X^i}{X^0}.
\end{equation} 
The first derivative of the prepotential with respect to $X^\Lambda$ is denoted by
\begin{equation}
	F_\Lambda = \frac{\pd F}{\pd X^\Lambda}.
\end{equation} 
Finally it makes sense to regroup the gauge fields into one single vector
\begin{equation}
	A^\Lambda = (A^0, A^i).
\end{equation} 

One needs to introduce two more quantities, respectively the Kähler potential and the Kähler connection
\begin{equation}
	K = - \ln i (\bar X^\Lambda F_\Lambda - X^\Lambda \bar F^\Lambda), \qquad
	\mc A_\mu = - \frac{i}{2} (\pd_i K\, \pd_\mu z^i - \pd_{\bar\imath} K\, \pd_\mu z^{\bar\imath}).
\end{equation} 

The Lagrangian of this theory is given by
\begin{equation}
	\mc L = - \frac{R}{2}
		+ g_{i\bar\jmath}(z, \bar z)\,z \pd_\mu z^i \pd^\nu z^{\bar\imath}
		+ \mc R_{\Lambda\Sigma}(z, \bar z)\, F^\Lambda_{\mu\nu} F^{\Sigma\,\mu\nu}
		- \mc I_{\Lambda\Sigma}(z, \bar z)\, F^\Lambda_{\mu\nu} \hodge{F}^{\Sigma\,\mu\nu}
\end{equation} 
where $R$ is the Ricci scalar and $\hodge{F}^\Lambda$ is the Hodge dual of $F^\Lambda$. The matrix
\begin{equation}
	\mc N = \mc R + i\, \mc I
\end{equation} 
can be expressed in terms of $F$. From this Lagrangian one can introduce the symplectic dual of $F^\Lambda$
\begin{equation}
	G_\Lambda = \frac{\var \mc L}{\var F^\Lambda} = \mc R_{\Lambda\Sigma} F^\Sigma - \mc I_{\Lambda\Sigma} \hodge{F}^\Sigma.
\end{equation} 

\subsection{BPS solutions}
\label{sec:N=2-ungauged-sugra:bps-solutions}

A BPS solution is a classical solution which preserves a part of the supersymmetry.
The BPS equations are obtained by setting to zero the variations of the fermionic partners under a supersymmetric transformation.
These equations are first order and under some conditions their solutions also solve the equations of motion.

In~\cite[sec.~3.1]{behrndt_stationary_1998} (see also~\cite[sec.~2.2]{hristov_bps_2010} for a summary), Behrndt, Lüst and Sabra obtained the most general stationary BPS solution for $N = 2$ ungauged supergravity.
The metric for this class of solutions reads
\begin{equation}
	\label{metric:sugra:static-N2}
	\dd s^2 = f^{-1} (\dd t + \omega\, \dd\phi)^2 + f\, \dd \Sigma^2,
\end{equation} 
with the $3$-dimensional spatial metric given in spherical or spheroidal coordinates
\begin{subequations}
\label{metric:flat-spatial}
\begin{align}
	\dd \Sigma^2 &= h_{ij}\, \dd x^i \dd x^j
		= \dd r^2 + r^2 (\dd\theta^2 + \sin^2 \theta\, \dd\phi^2) \\
		&= \dd r^2 + r^2 \dd\Omega^2 = \frac{\rho^2}{r^2 + a^2}\; \dd r^2 + \rho^2 \dd\theta^2 + (r^2 + a^2) \sin^2 \theta\; \dd \phi^2,
\end{align}
\end{subequations}
where $i, j, k$ are flat spatial indices (which should not be confused with the indices of the scalar fields).
The functions $f$ and $\omega$ depend on $r$ and $\theta$ only.

Then the solution is entirely given in terms of two sets of (real) harmonic functions~\footnote{We omit the tilde that is present in~\cite{behrndt_stationary_1998} to avoid the confusion with the quantities that are transformed by the JNA. No confusion is possible since the index position will always indicate which function we are using.} $\{ H^\Lambda, H_\Lambda \}$
\begin{subequations}
\label{eq:N=2-bps-equations}
\begin{gather}
	f = \e^{-K} = i (\bar X^\Lambda F_\Lambda - X^\Lambda \bar F_\Lambda), \\
	\levi{_{ijk}} \pd_j \omega_k = 2 \e^{-K} \mc A_i = (H_\Lambda \pd_i H^\Lambda - H^\Lambda \pd_i H_\Lambda), \\
	F^\Lambda_{ij} = \frac{1}{2}\, \levi{_{ijk}} \pd_k H^\Lambda, \qquad
	G_{\Lambda\,ij} = \frac{1}{2}\, \levi{_{ijk}} \pd_k H_\Lambda, \\
	i (X^\Lambda - \bar X^\Lambda) = H^\Lambda, \qquad
	i (F_\Lambda - \bar F_\Lambda) = H_\Lambda
\end{gather}
\end{subequations}
The only non-vanishing component of $\omega_i$ is $\omega \equiv \omega_\phi$.

Starting from the metric \eqref{metric:sugra:static-N2} in spherical coordinates with $\omega = 0$, one can use the JN algorithm of section~\ref{sec:janis-newman} with
\begin{equation}
	f_t = f^{-1}, \qquad
	f_r = f, \qquad
	f_\Omega = r^2 f
\end{equation} 
in order to obtain the metric \eqref{metric:sugra:static-N2} in spheroidal coordinates with $\omega \neq 0$ given by
\begin{equation}
	\omega = a (1 - \tilde f) \sin^2 \theta.
\end{equation} 
Then one needs only to find the complexification of $f$ and to check that it gives the correct $\omega$, as it would be found from the equations \eqref{eq:N=2-bps-equations}.
However it appears that one cannot complexify directly $f$.
Therefore one needs to complexify first the harmonic functions $H_\Lambda$ and $H^\Lambda$ (or equivalently $X^\Lambda$), and then to reconstruct the other quantities.
Nonetheless, equations \eqref{eq:N=2-bps-equations} ensure that finding the correct harmonic functions gives a solution, thus it is not necessary to check these equations for all the other quantities.

In the next subsections we provide two examples~\footnote{They correspond to singular solutions, but we are not concerned with regularity here.}, one for pure supergravity as an appetizer, and then one with $n_v = 3$ multiplets (STU model).

\subsubsection{Pure supergravity}

As a first example we consider pure (or minimal) supergravity, i.e. $n_v = 0$~\cite[sec.~4.2]{behrndt_stationary_1998}. The prepotential reads
\begin{equation}
	F = - \frac{i}{4}\, (X^0)^2.
\end{equation} 
The function $H_0$ and $H^0$ are related to the real and imaginary parts of the scalar $X^0$
\begin{equation}
	H_0 = \frac{1}{2} (X^0 + \bar X^0) = \Re X^0, \qquad
	\bar H^0 = i (X^0 - \bar X^0) = - 2 \Im X^0,
\end{equation} 
while the Kähler potential is given by
\begin{equation}
	f = \e^{-K} = X^0 \bar X^0.
\end{equation} 

The static solution corresponds to~\cite[sec.~4.2]{behrndt_stationary_1998}
\begin{equation}
	\label{eq:pure-sugra-static-X0}
	H_0 = X^0 = 1 + \frac{m}{r}
\end{equation} 
Performing the JN transformation with the rule \eqref{eq:scalar-complexification-rule} gives
\begin{equation}
	\tilde X^0 = 1 + \frac{m (r + i a \cos\theta)}{\rho^2}.
\end{equation}
This corresponds to the second solution of~\cite[sec.~4.2]{behrndt_stationary_1998} which is stationary with
\begin{equation}
	\omega = \frac{m (2r + m)}{\rho^2}\; a \sin^2 \theta.
\end{equation} 

\subsubsection{STU model}

We now consider the STU model $n_v = 3$ with prepotential~\cite[sec.~3]{behrndt_stationary_1998}
\begin{equation}
	F = - \frac{X^1 X^2 X^3}{X^0}.
\end{equation} 
The expressions for the Kähler potential and the scalar fields in terms of the harmonic functions are complicated and will not be needed; the curious reader can look at~\cite[sec.~3]{behrndt_stationary_1998}. Various choices for the functions will give different solutions.

A class of static black hole-like solutions are given by the harmonic functions~\cite[sec.~4.4]{behrndt_stationary_1998}
\begin{equation}
	\label{eq:stu-static-functions}
	H_0 = h_0 + \frac{q_0}{r}, \qquad
	H^i = h^i + \frac{p^i}{r}, \qquad
	H^0 = H_i = 0.
\end{equation} 
These solutions carry three magnetic $p^i$ and one electric $q_0$ charges.

Let's form the complex harmonic functions
\begin{equation}
	\mc H_0 = H_0 + i\, H^0, \qquad
	\mc H_i = H^i + i\, H_i.
\end{equation} 
Then the rule \eqref{eq:scalar-complexification-rule} leads to
\begin{equation}
	\mc H_0 = h_0 + \frac{q_0 (r + i a \cos\theta)}{\rho^2}, \qquad
	\mc H_i = h^i + \frac{p^i (r + i a \cos\theta)}{\rho^2},
\end{equation} 
for which the various harmonic functions read explicitly
\begin{equation}
	H_0 = h_0 + \frac{q_0 r}{\rho^2}, \qquad
	H^i = h^i + \frac{p^i r}{\rho^2}, \qquad
	H^0 = \frac{q_0 a \cos\theta}{\rho^2}, \qquad
	H_i = \frac{p^i a \cos\theta}{\rho^2}.
\end{equation}
This set of functions corresponds to the stationary solution of~\cite[sec.~4.4]{behrndt_stationary_1998} where the magnetic and electric dipole momenta are not independent parameters but obtained from the magnetic and electric charges instead.

\subsection{Dilaton–axion black hole – \texorpdfstring{$T^3$}{T3} model}

Sen derived his solution using the fact that Einstein–Maxwell gravity coupled to an axion $\sigma$ and a dilaton $\phi$ (for a specific value of dilaton coupling constant) can be embedded in heterotic string theory.
This model can also be embedded in $N = 2$ ungauged supergravity with $n_v = 1$, equal gauge fields $A \equiv A^0 = A^1$ and prepotential~\footnote{This model can be obtained from the STU model by setting the sections pairwise equal $X^2 = X^0$ and $X^3 = X^1$~\cite{chow_black_2014}.}
\begin{equation}
	F = - i\, X^0 X^1,
\end{equation} 
The dilaton and the axion corresponds to the complex scalar field
\begin{equation}
	z = \e^{-2\phi} + i\, \sigma.
\end{equation} 

The static metric, gauge field and the complex field read respectively
\begin{subequations}
\begin{align}
	\dd s^2 &= - \frac{f_1}{f_2}\, \dd t^2 + f_2 \Big(f_1^{-1}\, \dd r^2 + r^2\, \dd\Omega^2 \Big), \\
	A &= \frac{f_A}{f_2}\, \dd t, \\
	z &= \e^{-2\phi} = f_2
\end{align}
\end{subequations}
where
\begin{equation}
	f_1 = 1 - \frac{r_1}{r}, \qquad
	f_2 = 1 + \frac{r_2}{r}, \qquad
	f_A = \frac{q}{r}.
\end{equation} 
The radii $r_1$ and $r_2$ are related to the mass and the charge by
\begin{equation}
	r_1 + r_2 = 2 m, \qquad
	r_2 = \frac{q^2}{m}.
\end{equation} 

Applying now the Janis–Newman algorithm, the two functions $f_1$ and $f_2$ are complexified with the usual rules \eqref{eq:complexification-rules-1/r}
\begin{equation}
	\tilde f_1 = 1 - \frac{r_1 r}{\rho^2}, \qquad
	\tilde f_2 = 1 + \frac{r_2 r}{\rho^2}.
\end{equation} 
The final metric in BL coordinates is given by
\begin{equation}
	\dd s^2 = - \frac{\tilde f_1}{\tilde f_2} \left[ \dd t - a \left(1 - \frac{\tilde f_2}{\tilde f_1}\right) \sin^2 \theta\, \dd\phi \right]^2
		+ \tilde f_2 \left( \frac{\rho^2 \dd r^2}{\Delta} + \rho^2 \dd\theta^2 + \frac{\Delta}{\tilde f_1}\, \sin^2 \theta\, \dd\phi^2 \right)
\end{equation}
for which the BL functions \eqref{change:bl-gh} are
\begin{equation}
	g(r) = \frac{\hat \Delta}{\Delta}, \qquad
	h(r) = \frac{a}{\Delta}
\end{equation} 
with
\begin{equation}
	\label{eq:sen-bh-delta}
	\Delta = \tilde f_1 \rho^2 + a^2 \sin^2 \theta, \qquad
	\hat \Delta = \tilde f_2 \rho^2 + a^2 \sin^2 \theta.
\end{equation} 

Once $f_A$ has been complexified as
\begin{equation}
	\tilde f_A = \frac{q r}{\rho^2}
\end{equation} 
the transformation of the gauge field is straightforward
\begin{equation}
	A = \frac{\tilde f_A}{\tilde f_2}\, (\dd t - a \sin^2 \theta\, \dd\phi )
		- \frac{q r}{\Delta}\, \dd r.
\end{equation} 
The $A_r$ depending solely on $r$ can again be removed thanks to a gauge transformation.

One cannot complexify the scalar $z$ using the previous function $\tilde f_2$ since the latter is real and not complex.
Instead one needs to follow the rule \eqref{eq:scalar-complexification-rule} a new time in order to obtain
\begin{equation}
	z = 1 + \frac{r_2}{r} = 1 + \frac{r_2 \bar r}{\rho^2}.
\end{equation} 
The explicit values for the dilaton and axion are then
\begin{equation}
	\e^{-2\phi} = \tilde f_2, \qquad
	\sigma = \frac{r_2 a \cos \theta}{\rho^2}.
\end{equation} 

We have been able to find the full Sen's solution, completing the computations from~\cite{yazadjiev_newman-janis_2000}.
It is interesting to note that for another value of the dilaton coupling we cannot use the transformation~\footnote{The authors of~\cite{hansen_applicability_2013} report incorrectly that~\cite{pirogov_towards_2013} is excluding all dilatonic solutions.}~\cite{horne_rotating_1992, pirogov_towards_2013}.
Finally the truncation $\sigma = 0$ is also a solution of dilatonic gravity~\cite{horne_rotating_1992}, but the JN algorithm generates directly the axion–dilaton metric such that we can not recover the vanishing axion case~\cite{yazadjiev_newman-janis_2000}.

\section{Conclusion and perspectives}

In this paper we showed how to apply the DJN formalism to solution with complex parameters and complex scalar fields through several examples.
From our results the picture of the DJN algorithm is now (almost) complete since we gave the rules for all possible bosonic fields and for all usual parameters.
Namely, the last missing piece would be the inclusion of acceleration $\alpha$, which is the last parameter of Plebański-Demiański class of solutions~\cite{plebanski_class_1975, plebanski_rotating_1976}, parametrized by the six parameters $(m, n, q, p, a, \alpha)$.

It is to notice, that the appearance of complex coordinate transformations mixed with complex parameter transformations was a feature of Quevedo's solution generating technique~\cite{quevedo_complex_1992, quevedo_determination_1992}.
Yet it is unclear what the link with our approach really is, despite the fact that it may probably provide some clues for generalizing further the DJN algorithm (higher dimensions, cosmological backgrounds…).

Another interesting point is that all the examples presented in this paper are truncations of the Chow–Compère black hole~\cite{chow_black_2014}, and it would be useful to understand in which cases the DJN algorithm can be applied to this solution.

A further step could also be to apply this new formalism to gauged supergravity.
In this case the cosmological constant is typically non-vanishing which would imply that one can add only the NUT charge~\cite{demianski_new_1972}, and the complexification of the mass is a more complicated~\cite{erbin_deciphering_2014}.

\section*{Acknowledgments}

We want to thank N. Halmagyi for valuable comments on an earlier version of this work.
H. E. is also grateful to G. Compère and D. Klemm for interesting discussions.
L. H. would like to thank the DESY theory group of Hamburg and the GGI institute (Firenze) for hospitality during the realization of this work.

For this work, made within the \textsc{Labex Ilp} (reference \textsc{Anr–10–Labx–63}), H. E. was supported by French state funds managed by the \emph{Agence nationale de la recherche}, as part of the programme \emph{Investissements d'Avenir} under the reference \textsc{Anr–11–Idex–0004–02}.

\appendix

\section{Supergravity solutions with a NUT charge}
\label{app:sugra-nut}

\subsection{Pure supergravity}

In~\cite[sec.~4.2]{behrndt_stationary_1998} a solution of pure supergravity (see~\ref{sec:N=2-ungauged-sugra:bps-solutions} for the notations) with a NUT charge is presented.
In this case the solution reads
\begin{equation}
	\label{eq:pure-sugra-static-X0-nut}
	X^0 = 1 + \frac{m + i n}{r}, \qquad
	\omega = 2n \cos\theta.
\end{equation} 

The question is whether this configuration can be obtained from the $n = 0$ solution \eqref{eq:pure-sugra-static-X0}
\begin{equation}
	X^0 = 1 + \frac{m}{r}
\end{equation} 
from the transformation \eqref{eq:complex-mass-no-lambda}
\begin{equation}
	m = m' + i n, \qquad
	r = r' + i n.
\end{equation} 
It is straightforward to check that the full metric \eqref{metric:sugra:static-N2} is recovered, while the field $X^0$ in \eqref{eq:pure-sugra-static-X0-nut} follows from the rule \eqref{eq:complexification-rules-r}
\begin{equation}
	r \longrightarrow \frac{1}{2}\, (r + \bar r) = \Re r = r'
\end{equation} 
applied in the denominator.
Hence a DJN transformation with the NUT charge does not act in the same way as a transformation with an angular momentum, since the transformation rule is different from \eqref{eq:scalar-complexification-rule}.

\subsection{SWIP solutions}

Let's consider the action~\cites{bergshoeff_stationary_1996}[sec.~12.2]{ortin_gravity_2004}
\begin{equation}
	S = \frac{1}{16 \pi} \int \dd^4 x\, \sqrt{\abs{g}}\, \left( R
		- 2 (\pd \phi)^2 - \frac{1}{2}\, \e^{4\phi}\, (\pd \sigma)^2
		- \e^{-2\phi} F^i_{\mu\nu} F^{i\mu\nu} + \sigma\, F^i_{\mu\nu} \tilde{F}^{i\mu\nu} \right)
\end{equation} 
where $i = 1, \ldots, M$.
When $M = 2$ and $M = 6$ this action corresponds respectively to $N = 2$ supergravity with one vector multiplet and to $N = 4$ pure supergravity, but we keep $M$ arbitrary.
The axion $\sigma$ and the dilaton $\phi$ are naturally paired into a complex scalar
\begin{equation}
	z = \sigma + i \e^{-2\phi}.
\end{equation} 

In order to avoid redundancy we first provide the general metric with $a, n \neq 0$, and we explain how to find it from the restricted case $a = n = 0$.

Stationary Israel–Wilson–Perjés (SWIP) solutions correspond to
\begin{subequations}
\label{eq:solution-swip}
\begin{gather}
	\dd s^2 = - \e^{2U} W (\dd t + A_\phi\, \dd \phi)^2 + \e^{-2U} W^{-1} \dd \Sigma^2, \\
	A^i_t = 2 \e^{2U} \Re(k^i H_2), \qquad
	\tilde A^i_t = 2 \e^{2U} \Re(k^i H_1), \qquad
	z = \frac{H_1}{H_2}, \\
	A_\phi = 2 n \cos \theta - a \sin^2 \theta (\e^{-2U} W^{-1} - 1), \\
	\e^{-2U} = 2 \Im(H_1 \bar H_2), \qquad
	W = 1 - \frac{r_0^2}{\rho^2}.
\end{gather}
\end{subequations}
This solution is entirely determined by the two harmonic functions
\begin{equation}
	\label{eq:swip:harmonic-functions}
	H_1 = \frac{1}{\sqrt{2}}\, \e^{\phi_0} \left( z_0 + \frac{z_0 \mc M + \bar z_0 \Upsilon}{r - i a \cos \theta} \right), \qquad
	H_2 = \frac{1}{\sqrt{2}}\, \e^{\phi_0} \left( 1 + \frac{\mc M + \Upsilon}{r - i a \cos \theta} \right).
\end{equation} 
The spatial $3$-dimensional metric $\dd \Sigma^2$ reads
\begin{equation}
	\label{metric:flat-spatial-swip}
	\dd\Sigma^2 = h_{ij}\, \dd x^i \dd x^j
		= \frac{\rho^2 - r_0^2}{r^2 + a^2 - r_0^2}\; \dd r^2 + (\rho^2 - r_0^2) \dd\theta^2 + (r^2 + a^2 - r_0^2) \sin^2 \theta\; \dd \phi^2.
\end{equation} 

Finally, $r_0$ corresponds to
\begin{equation}
	r_0^2 = \abs{\mc M}^2 + \abs{\Upsilon}^2 - \sum_i \abs{\Gamma^i}^2
\end{equation} 
where the complex parameters are
\begin{equation}
	\mc M = m + i n, \qquad
	\Gamma^i = q^i + i p^i,
\end{equation} 
$m$ being the mass, $n$ the NUT charge, $q^i$ the electric charges and $p^i$ the magnetic charges, while the axion/dilaton charge $\Upsilon$ takes the form
\begin{equation}
	\Upsilon = - \frac{1}{2} \sum_i \frac{(\bar \Gamma^i)^2}{\mc M}.
\end{equation} 
The latter together with the asymptotic values $z_0$ comes from
\begin{equation}
	z \sim z_0 - i \e^{-2\phi_0} \frac{2 \Upsilon}{r}.
\end{equation} 
The complex constant $k^i$ are determined by
\begin{equation}
	k^i = - \frac{1}{\sqrt{2}}\, \frac{\mc M \Gamma^i + \bar \Upsilon \bar \Gamma^i}{\abs{\mc M}^2 - \abs{\Upsilon}^2}.
\end{equation} 

As discussed in the previous appendix, the transformation of scalars fields is different whether one is turning on a NUT charge or an angular momentum.
For this reason, starting from the case $a = n = 0$, one needs to perform the two successive transformations
\begin{subequations}
\begin{gather}
	\label{eq:swip:djn-nut}
	u = u' - 2 i n \ln \sin \theta, \qquad
	r = r' + i n, \qquad
	m = m' + i n, \\
	\label{eq:swip:djn-rot}
	u = u' + i a \cos \theta, \qquad
	r = r' - i a \cos \theta,
\end{gather}
\end{subequations}
the order being irrelevant (for definiteness we choose to add the NUT charge first).
As explained in~\cite{erbin_deciphering_2014}, group properties of the DJN algorithm ensure that the metric will be transformed as if only one transformation was performed, and one can use the formula of section~\ref{sec:janis-newman}.
Then the formulas \eqref{eq:stationary-metric:tr} for the metric and \eqref{eq:stationary-gauge-field:tr} for the gauge field directly apply, which ensures that the general form of the solution \eqref{eq:solution-swip} is correct~\footnote{For that one needs to shift $r^2$ by $r_0^2$ in order to bring the metric \eqref{metric:flat-spatial-swip} to the form \eqref{metric:flat-spatial}. This modifies the function but one does not need this fact to obtain the general form. Then one can shift by $- r_0^2$ before dealing with the complexification of the functions.}.
Since all the functions and the parameters depend only on $\mc M$, $H_1$ and $H_2$, it is sufficient to explain their complexification.

The function $W$ is easily transformed, whereas $H_1$ and $H_2$ are more subtle since they are complex harmonic functions.
Let's consider first the NUT charge with the transformation \eqref{eq:swip:djn-nut}. According to the previous appendix, the $r$ in the denominator of both functions is transformed according to \eqref{eq:complexification-rules-r}
\begin{equation}
	r \longrightarrow \frac{1}{2}\, (r + \bar r) = \Re r = r'.
\end{equation} 
Then one can perform the second transformation \eqref{eq:swip:djn-rot} in order to add the angular momentum.
Using the recipe from section~\ref{sec:N=2-ungauged-sugra:complex-rule}, one obtain the correct result \eqref{eq:swip:harmonic-functions} by just replacing $r$ with \eqref{eq:swip:djn-rot}.

Finally let's note that it seems possible to also start from $p^i = 0$ and to turn them on using the transformation
\begin{equation}
	q^i \longrightarrow q'^i = q^i + i p^i,
\end{equation} 
using different rules for complexifying the various terms (depending on the fact if one is dealing with a real or a complex function/parameter).

\printbibliography[heading=bibintoc]

\end{document}